\documentstyle[aps,pre,multicol,epsfig]{revtex}

\begin{document}
\draft

\title{Simplest random $K$-satisfiability problem}

\author{Federico Ricci-Tersenghi~\cite{frt}, Martin Weigt~\cite{mw},
and Riccardo Zecchina~\cite{rz}}

\address{
\cite{frt,rz} The Abdus Salam International Centre for
Theoretical Physics, Condensed Matter Group\\
Strada Costiera 11, P.O. Box 586, I-34100 Trieste, Italy\\
\cite{mw} Institute for Theoretical Physics, University of
G\"ottingen, Bunsenstr. 9, D-37073 G\"ottingen, Germany
}

\maketitle

\begin{abstract}
We study a simple and exactly solvable model for the generation of
random satisfiability problems. These consist of $\gamma N$ random
boolean constraints which are to be satisfied simultaneously by $N$
logical variables. In statistical-mechanics language, the considered
model can be seen as a diluted p-spin model at zero temperature. While
such problems become extraordinarily hard to solve by local search
methods in a large region of the parameter space, still at least one
solution may be superimposed by construction.  The statistical
properties of the model can be studied exactly by the replica method
and each single instance can be analyzed in polynomial time by a
simple global solution method.  The geometrical/topological structures
responsible for dynamic and static phase transitions as well as for
the onset of computational complexity in local search method are
thoroughly analyzed.  Numerical analysis on very large samples allows
for a precise characterization of the critical scaling behaviour.
\end{abstract}


\section{Introduction}

Complexity theory~\cite{NPC}, as arising from Cook's theorem of
1971~\cite{Cook}, deals with the issue of classifying combinatorial
optimization problems according to the computational cost required for
their solution.  The hard problems are grouped in a class named NP,
where NP stands for `non-deterministic polynomial time'.  These
problems are such that a potential solution can be checked rapidly
whereas finding one solution may require a time growing exponentially
with system size in worst case.  In turn, the hardest problems in NP
belong to a sub-class called NP-complete which is at the root of
computational complexity.  The completeness property refers to the
fact that if an efficient algorithm for solving just one of these
problems could be found, then one would have an efficient algorithm
for solving all problems in NP.  By now, a huge number of NP-complete
problems have been identified~\cite{NPC}, and the lack of an efficient
algorithm corroborates the widespread conjecture that no such
algorithm exists, or more formally that NP$\ne$P where P includes all
problems solvable in polynomial time.

Complexity theory is based on a worst-case analysis and therefore does
not depend on the properties of the particular instances of the
problems under consideration.  In order to deepen the understanding of
typical-case complexity rather than the worst-case one and to improve
and test algorithms for real world applications, computer scientists
have recently focused their attention on the study of random instances
of hard computational problems, seeking for a link between the onset
of computational complexity and some intrinsic (i.e.\ algorithm
independent) properties of the model. Analytical and numerical results
have accumulated~\cite{FuAnd,hard,KirkSel,AI,nature,martin} showing
that the computationally hard instances appear with a significant
probability only when generated near ``phase boundaries'', i.e.\ when
problems are critically constrained. This phenomenon is know as the
easy-hard transition.

Randomized search algorithms provide efficient heuristics for quickly
finding solutions provided they exist. At the phase boundary, however,
there appears an exponential critical slowing down which makes the
search inefficient for any practical purpose. Understanding the
behaviour of search processes at the easy-hard transition point
constitutes an important theoretical challenge which can be viewed as
the problem of building a generalized off-equilibrium theory for
stochastic processes which do not satisfy detailed balance.  No static
probability measure describing the asymptotic statistical behaviour of
the search processes is guaranteed to exist.  Moreover, the hardest
random instances of combinatorial optimization problems provide a
natural test-bed for the optimization of heuristic search algorithms
which are widely used in practice.

How to generate hard and solvable instances is far from obvious and
very few examples of such generators are known~\cite{HARD_SOLV}.  In
most cases, like e.g.\ in the random Boolean satisfiability problem
(K-SAT~\cite{AI,nature,chayes}), for a short definition see note
\cite{note2}, hard instances can be only found in a very narrow region
of the parameters space.  In this region the probability that a random
instance of the problem has no solution at all is finite.  Then,
heuristic (incomplete) search algorithms have no way to disentangle,
in a given finite time, the unsatisfiable instances, from those which
are simply very hard to solve.

In this paper, we shall discuss a very simple and exactly solvable
model for the generation of random combinatorial problems.  On one
hand, these become extraordinarily hard to solve by local search
methods in a large region of the parameter space and yet at least one
solution may be superimposed by construction.  On the other hand, the
model may be solved in polynomial time by a simple global method and
therefore belongs to the class P.

At variance with respect to the famous random 2-SAT
problem~\cite{goerdt,chayes}, which is in P and can be solved
efficiently by local search methods~\cite{algo} also at the phase
boundary, the model we consider undergoes an easy-hard transition very
similar (even harder) to the one observed in 3-SAT as far as local
search methods is concerned. However, the exact mapping of the model
on a minimization problem over uniform random hyper-graphs makes the
problem analytically tractable. It is also solvable in polynomial time
by a global method which allows for the numerical study of very large
systems. Therefore, some of the open questions which arise the the
analysis of 3-SAT and which are common to the present model can be
answered exactly.

In the context of statistical physics the model provides a simple
model for the glass transition, in which the crystalline state can be
view as the superimposed solution and the structure of the excited
states is responsible for the off-equilibrium behaviour and the
associated structural glass transition. These aspects will be the
subject of a forthcoming paper.  The limit of infinite connectivity
provides one of the most studied models in the context of spin glass
theory, see for instance~\cite{MPV,Gross,Gardner,GD,TT,BZ}.

\section{The model}
\label{sec:model}

In order to unveil the different aspects of the model, to be referred
to as {\it hyper-SAT} (hSAT), we give explicitly its definition both
as a satisfiability problem and as a minimization problem over
hyper-graphs.

Here we discuss the hSAT model with $K=3$ variables per constraint,
which can be viewed as a perfectly balanced version of the famous
random 3-SAT problem.  The case $K=2$ does not present any interesting
computational features as far as hardness is concerned because it can
be solved efficiently both by local and global
methods. Generalizations to $K>3$ are straightforward.

Given a set of $N$ Boolean variables $\{x_i=0,1\}_{i=1,\ldots,N}$, we
construct an instance of 3-hSAT as follows.  Firstly we define the
following elementary constraints (4-clauses sets with 50\% satisfying
assignments)
\begin{eqnarray}
C(ijk|+1) &=& (x_i \vee x_j \vee x_k) \wedge (x_i \vee \bar x_j \vee
\bar x_k) \wedge (\bar x_i \vee x_j \vee \bar x_k) \wedge (\bar x_i
\vee \bar x_j \vee x_k) \nonumber \\
C(ijk|-1) &=& (\bar x_i \vee \bar x_j \vee \bar x_k) \wedge (\bar x_i
\vee x_j \vee x_k) \wedge (x_i \vee \bar x_j \vee x_k) \wedge (x_i
\vee x_j \vee \bar x_k) \;\;,
\label{c4}
\end{eqnarray}
where $\wedge$ and $\vee$ stand for the logical AND and OR operations
respectively and the over-bar is the logical negation.  Then, by
randomly choosing a set $E$ of $M$ triples $\{i,j,k\}$ among the $N$
possible indices and $M$ associated unbiased and independent random
variables $J_{ijk}=\pm 1$, we construct a Boolean expression in
Conjunctive Normal Form (CNF) as
\begin{equation}
F = \bigwedge_{\{i,j,k\}\in E} C(ijk|J_{ijk}) \;\;.
\label{Fcnf}
\end{equation}
A logical assignment of the $\{x_i\}$s satisfying all clauses, that is
evaluating $F$ to true, is called a solution of the 3-hSAT problem. If
no such assignment exists, $F$ is said to be unsatisfiable.

A slightly different choice of $J_{ijk}$ allows to construct hSAT
formulas which are random but guaranteed to be satisfiable. To every
Boolean variable we associate independently drawn random variables
$\varepsilon_i=\pm 1$, and define
$J_{ijk}=\varepsilon_i\varepsilon_j\varepsilon_k$ for all
$\{i,j,k\}\in E$.  For this choice, CNF formula (\ref{Fcnf}) is
satisfied by $\{x_i\ |\ x_i=+1 \mbox{ if } \varepsilon_i=+1, \ x_i=0
\mbox{ if } \varepsilon_i=-1\}$.  As we shall discuss in great detail,
these formulas provide a uniform ensemble of hard satisfiable
instances for local search methods.  We refer to this version of the
model as the {\it satisfiable hSAT}. Indeed, the random signs of
$J_{ijk}$ can be removed in this satisfiable case by negating all
Boolean variables $x_i$ associated to negative $\varepsilon_i$. The
resulting model has $J_{ijk}=+1$ for all $\{i,j,k\}\in E$, and the
forced satisfying solution is $x_i=1,\ \forall i=1,...,N$. The use of
the $\{\varepsilon_i\}$ is a way of hiding the latter solution by a
{\it random gauge transformation} without changing the properties of
the model. The impossibility of inverting efficiently the gauge
transformation by local methods is a consequence of the branching
process arising form the presence of $K=3$ variables in each
constraint.  For any $K>3$ the same result would hold whereas for
$K=2$ the problem trivializes.

The hSAT model can be easily described as a minimization problem of a
cost-energy function over a random hyper-graph.  Given a random
hyper-graph ${\cal G}_{N,M}=(V,E)$, where $V$ is the set of $N$
vertices and $E$ is the set of $M$ hyper-edges joining triples of
vertices, the energy function to be minimized reads
\begin{equation}
H_J[{\bf S}] = M-\sum_{\{i,j,k\}\in E} J_{ijk} \, S_i S_j S_k \;\;,
\label{Ham}
\end{equation}
where each vertex $i$ bears a binary ``spin'' variable $S_i=\pm 1$,
and the weights $J_{ijk}$ associated to the random bonds can be either
$\pm 1$ at random, in the so called {\it frustrated} case, or simply
equal to $1$ in the {\it unfrustrated} model.

Once the mapping $S_i=1$ if $x_i=1$ and $S_i=-1$ if $x_i=0$ is
established, one can easily notice that the energy function in
Eq.(\ref{Ham}) simply counts twice the number of violated clauses in
the previously defined CNF formulas with the same set of $J$'s.  The
frustrated and the unfrustrated cases correspond to the hSAT and to
the satisfiable hSAT formulas respectively.

The computational issue consists in finding a configuration of spin
variables which minimizes $H$. If all the terms $J_{ijk} S_i S_j S_k$
appearing in the energy are simultaneously maximized (``satisfied'')
the energy vanishes. This is always possible in the unfrustrated case
just by setting $S_i=1$, $\forall i$. In the frustrated case, there
exist a critical value of the average connectivity above which the
various terms start to be in conflict, that is frustration becomes
effective in the model.  In random hyper-graphs the control parameter
is the average density of bonds, $\gamma=M/N$ (or, for the CNF
formula, the density of clauses $\alpha= 4 \gamma$). For sufficiently
small densities, the graph consists of many small connected clusters
of size up to $O(\ln N)$. If $\gamma$ increases up to the percolation
value $\gamma_p=1/6$, there appears a giant cluster containing a
finite fraction of the $N$ sites in the limit of large $N$. However,
also this cluster can a priori have a tree-like structure, for which
the randomness of the couplings $J_{ijk}=\pm 1$ can be eliminated by a
proper gauge transformation, $S_i \to \pm S_i'$, of the spin
variables. As we shall see, there exist two other thresholds of the
bond density at which more complicated and interesting dynamical and
structural changes take place.

In spite of apparent similarities, hSAT and the random Boolean
Satisfiability problem (K-SAT~\cite{note2}) differ in some basic
aspects.

In K-SAT the fluctuations of the frequencies of appearance of the
variables in the clauses lead to both single and two body interactions
in the associated energy function~\cite{note3} which force the minima
in some specific random directions and which rule out the existence of
a purely dynamical threshold (see below).  Algorithms may take
advantage of such information and both heuristic as well as complete
algorithms show a performance which indeed depends on the criterion
used to fix the variables. For example, rigorous lower bounds to the
critical threshold have been improved recently by exploiting this
opportunity in a simple tractable way~\cite{Achlioptas}.  On the same
footing, the efficiency of the most popular heuristic and complete
search algorithms, namely walk-sat~\cite{wsat} and
TABLEAU~\cite{tableau}, is again based on strategies which exploit the
above structure.  Note that the above improvements cannot be applied
to the hSAT model where formulas are completely balanced.

Moreover, in K-SAT the mapping of the problem over directed random
graphs is rather involved and the exact analytical solution is still
lacking, while in hSAT the connection to random hyper-graphs is clear
and makes the analysis tractable.

Finally, restriction of K-SAT to satisfiable instances (for instance
by selecting at random clauses which are satisfied by a previously
fixed assignment of variables) does not provide a uniform ensemble of
hard satisfiable problems even when restricted to local search
methods~\cite{note4}.

Given the mapping over random hyper-graphs, the satisfiability problem
for hSAT can be solved in $O(N^3)$ steps by simply noticing that the
problem of satisfying all constrains is nothing but the problem of
solving an associated linear system modulo 2, i.e.\ in $GF[2]$. Upon
introducing the two sets of binary variables $\{a_i\} \in \{0,1\}^N$
and $\{b_{ijk}\} \in \{0,1\}^M$ such that $(-1)^{a_i}=S_i$ and
$(-1)^{b_{ijk}}=J_{ijk}$, the hSAT decision problem becomes simply the
problem of determining the existence of a solution in $GF[2]$ to the
random linear system $a_i+a_j+a_k=b_{ijk} ({\rm mod}\ 2)$, with ${ijk}$
running over all triples.

Finally, we notice that in the high $\gamma$ UNSAT (or frustrated)
region the optimization problem of minimizing the number of violated
constrains, the so called MAX-hSAT, is indeed computationally very
hard both for complete and incomplete algorithms and no global method
for finding ground states is available.

\section{Outline of main results}

For the sake of clarity, we anticipate here the main results leaving
for the following sections a thorough discussion of the analytical and
numerical studies.

The frustrated hSAT model presents two clear transitions.  The first
one appears at $\gamma_d = 0.818$ and it is of purely dynamical
nature. There the typical formula still remains satisfiable with
probability one, but an exponential number of local energy minima
appear at positive energies.  Deterministic algorithms, like greedy
search or zero temperature dynamics, try to decrease the energy in
every step and thus get stuck at least at this threshold. Randomized
algorithms may escape from these minima, but they undergo a slowing
down from an exponentially fast convergence to a polynomially slow
one, i.e.\ at $\gamma_d$ the typical time for finding a solution
diverges as a power of the number of variables. The dynamical
transition at $\gamma_d$ seems to be accompanied by a dynamical glassy
transition due to replica symmetry breaking (RSB) effects connected
with the appearance of an exponential number of local minima. An
approximate variational calculation (see ref.~\cite{variational} for a
discussion on the method) involving RSB gives $\gamma_{d,rsb}^{var}
\simeq 0.83$ which is in good agreement with the value of $\gamma_d$
where local minima appear.

The second transition appears at $\gamma_c= 0.918$ and corresponds to
the so-called SAT/UNSAT transition (below $\gamma_c$ the typical
problem is satisfiable whereas above $\gamma_c$ it becomes
unsatisfiable).  At this point the structure of the global energy
minima changes abruptly. The ground states have strictly positive
energy, thus no satisfying assignments for the hSAT formula exist any
more.  While the number of these configurations is always
exponentially large (the ground state entropy is always finite), at
$\gamma_c$ a finite fraction of the variables, the so-called {\it
backbone} component, becomes totally constraint, i.e.\ the backbone
variables take the same value in all minima~\cite{note9}. An important
difference of the SAT/UNSAT transition in hSAT compared to
K-SAT~\cite{variational} is the non-existence of any precursor.  For
$\gamma<\gamma_c$ and large $N$, all variables $S_i$ take equally
often the values +1 and $-1$ in the ground states (they have zero
local magnetization), even those which become backbone elements when
$\gamma_c$ is reached by adding new 4-clauses sets.  The lack of any
precursor comes from the non-existence of single- or two-body
interactions in Eq.(\ref{Ham}).

The unfrustrated or satisfiable hSAT problem has by construction at
least one solution which we find to be superimposed without affecting
the statistical features of the model for $\gamma<\gamma_c$ in the
limit of large $N$, including the dynamical transition at $\gamma_d=
0.818$. It is impossible to get any information on the superimposed
solution by looking at the full solution space, because it is
completely hidden by the exponential number of ground states. Randomly
chosen satisfying assignments do not show any correlation. At exactly
the same $\gamma_c= 0.918$ as in the frustrated model, there appears a
transition from a SAT phase with exponentially many unbiased solutions
to another SAT phase where the solutions are strongly concentrated
around the superimposed solution. The latter one is now hidden by the
presence of exponentially many local energy minima with positive
cost. These minima have exactly the statistical properties of the
global minima of the corresponding frustrated hSAT problem, that is
the hSAT problem defined over the same hyper-graph but with randomized
signs of the couplings $J_{ijk}$.  More specifically, the energy, the
entropy and the backbone component size coincide. Due to their finite
entropy, an algorithm will thus hit many of these local minima before
it reaches the satisfying ground state. As one can see from
Fig.~\ref{FIG5}, finding this solution by backtracking, e.g.\ with the
Davis-Putnam (DP) procedure~\cite{DP}, is nevertheless easier than
proving the unsatisfiability of hSAT (or identifying ground states in
the frustrated version). This results stems from the missing
information on the true ground state energy of hSAT above
$\gamma_c$. The solution time is however found to be clearly
exponential in both cases.  In the $\gamma \ge\gamma_c$ region, the
model provides a uniform ensemble of hard SAT instances for local
methods which can be used to test and optimize algorithms.

\section{Statistical mechanics analysis: the replica results}

In our analytical approach, we exploit the well known analogies
between combinatorial optimization problems and statistical
mechanics. In both cases, the system is characterized by some
cost-energy function, as it is given e.g.\ by Eq.(\ref{Ham}) for hSAT.
In equilibrium statistical mechanics, any configuration ${\bf
S}=\{S_i\}_{i=1...N}$ is realized with probability $\exp\{- \beta
H[{\bf S}] \}/Z$ where $\beta=1/T$ is the inverse temperature and $Z$
the partition function.  If the temperature is lowered, the
probability becomes more and more concentrated on the global energy
minima and finally, for $T=0$, only the ground states keep non-zero
weights.  

In order to compute the average free energy, we resort to
the replica symmetric (RS) functional replica method developed for
diluted spin glasses which is known to provide exact results for
ferromagnetic models: To circumvent the difficulty of computing the
average value of $\ln Z$, we compute the
$n^{th}$ moment of $Z$ for integer-valued $n$ and perform an
analytical continuation to real $n$ to exploit the identity $\ln \ll
Z^n \gg = 1 + n \ll \ln Z \gg + O(n^2)$.  The $n^{th}$ moment of $Z$
is obtained by replicating $n$ times the sum over the spin
configuration and averaging over the disorder~\cite{MZ}
\begin{equation}
\ll Z^n \gg = \sum_{{\bf S}^1 , {\bf S}^2 , \ldots ,{\bf S}^n} \ll
\exp \left( - \beta \sum_{a=1}^n H_J[{\bf S}^a] \right) \gg \quad ,
\label{nthmoment}
\end{equation}
which in turn may be viewed as a generating function in the variable
$\exp(-\beta)$.

In order to compute the expectation values that appear in
Eq.(\ref{nthmoment}), one notices that each single term $ \exp (
-\beta \sum _{a=1} ^n H_J[{\bf S}^a ] )$ factorises over the sets of
different triples of indices due to the absence of any correlation in
the probability distribution of the $J_{ijk}$. It follows
\begin{equation}
\ll Z ^ n \gg = \sum _{S_i^1 , S_i^2 , \ldots,S_i^n } \exp \left
\{ - \beta \gamma N n-\gamma N+\frac{\gamma}{N^2} \sum_{ijk} e^{\beta
\sum_a S_i^a S_j^a S_k^a} +O(1) \right \}
\label{nthmoment2}
\end{equation}
The averaged term in Eq.(\ref{nthmoment2}) depends on the $n\times N$
spins only through the $2^n$ occupation fractions $c(\vec \sigma)$
labelled by the vectors $\vec \sigma$ with $n$ binary components;
$c({\vec \sigma})N$ equals the number of labels $i$ such that
$S_i^a=\sigma ^a$, $\forall a=1,\ldots ,n$. Therefore, the final
expression of the $n^{th}$ moment of $Z$ to the leading order in $N$
(i.e.\ by resorting to a saddle point integration), can be written as
$\ll Z^n \gg \simeq \exp ( N \; F[c] )$ where $F[c]$ is the maximum
over all possible $c(\vec\sigma)$'s of the functional~\cite{MZ}
\begin{equation}
-\beta F[c] = -\gamma (1+\beta n)-\sum_{\vec \sigma} c(\vec \sigma)
 \ln c(\vec \sigma)+\gamma \sum_{\vec \sigma,\vec \rho,\vec \tau}
 c(\vec \sigma) c(\vec \rho) c(\vec \tau) \exp(\beta \sum_a \sigma^a
 \rho^a \tau^a) \; \; \; .
\label{free}
\end{equation}
The saddle point equation $\frac{\partial (-\beta F)}{\partial c(\vec
\sigma)} = \Lambda - 1$ reads
\begin{equation}
c(\vec \sigma)=\exp \left \{-\Lambda+3 \gamma \sum_{\vec \rho,\vec
\tau} c(\vec \rho) c(\vec \tau) \exp(\beta \sum_a \sigma^a \rho^a
\tau^a) \right \}
\label{saddle}
\end{equation}
where the Lagrange multiplier $\Lambda$ enforces the
normalization constraint $\sum_{\vec \sigma} c(\vec \sigma)=1$, and
goes to $3\gamma$ for $n\to 0$.  In
Eq.(\ref{free}), one may easily identify two terms, one model
dependent and the other ($ - \sum _{\vec \sigma} c({\vec \sigma}) \ln
c({\vec \sigma})$) simply describing the degeneracy (the so called
combinatorial {\it entropy}) with which each term of the generating
function appears given the representation in terms of the occupation
fractions.  In the limit of interest $T \to 0$ and in the replica
symmetric subspace, the freezing of the spin variables is properly
described by a rescaling of the local magnetizations of the form
$m=\tanh (\beta h)$.  The probability distribution $P(h)$ is
therefore introduced through the generating functional
\begin{equation}
c(\vec \sigma) =\int_{-\infty}^{\infty}dh\ P(h) \frac{e^{\beta h 
\sum_a \sigma^a}}{(2 \cosh(\beta h))^n}
\label{ph}
\end{equation}
where $h$ is nothing but an effective field in which the spins are
immersed. $c$ depends on $\vec\sigma$ only via $s=\sum_a\sigma^a$. 
In this representation, the free-energy reads
\begin{eqnarray}
&-&\beta F[P(h)] = \nonumber \\
&\gamma&  \int dh_1 dh_2 dh_3 P(h_1) P(h_2) P(h_3)
\ln\frac{2 \cosh(\beta (h_1+h_2)) [e^{\beta h_3}+e^{-2 \beta-\beta
h_3}]+2 \cosh (\beta (h_1-h_2)) [e^{-\beta h_3}+e^{-2 \beta+\beta
h_3}]}{(2 \cosh (\beta h_1))(2 \cosh (\beta h_2))(2 \cosh(\beta
h_3))} \nonumber \\
&+& \int \frac{dh\,dK}{2 \pi} e^{i h K} P_{FT}(K)[1-\ln P_{FT}(K)]
\ln[2 \cosh(\beta h)]
\label{free2}
\end{eqnarray}
where $P_{FT}(K)$ is the Fourier transform of $P(h)$.  The associated
saddle-point equation reads
\begin{equation}
\int dh\,P(h)\,e^{\beta h s}=\exp\left \{ -3 \gamma+ 3
\gamma \int dh_1 dh_2 P(h_1) P(h_2) G(h_1,h_2) \right \} \; \; ,
\label{saddle2}
\end{equation}
where
\begin{equation}
G(h_1,h_2)=\left(
\frac{\cosh(\beta (h_1+h_2))+e^{-2 \beta}\cosh(\beta (h_1-h_2))}
{\cosh(\beta (h_1-h_2))+e^{-2 \beta}\cosh(\beta (h_1+h_2))}
\right)^{\frac{s}{2}} \; \; \;.
\end{equation}

In the case of {\it satisfiable hSAT}, at $\beta \to \infty$ ($T=0$)
and in the version having no random gauge ($J_{ijk}=+1, \forall
\{i,j,k\}\in E$), the spins turn out to be subject to an effective
local field $h$ which fluctuates from site to site according to the
following simple probability distribution
\begin{equation}
 P(h)=\sum_{\ell \geq 0} p_{\gamma}^{(\ell)} \delta(h-\ell)
\end{equation}
with the saddle-point conditions
\begin{eqnarray}
\label{eq:psaddle}
p_{\gamma}^{(\ell)}&=&(3 \gamma)^\ell \frac{(1-p_{\gamma}^{(0)})^{2
\ell} p_{\gamma}^{(0)}} {\ell ! }\nonumber\\
p_\gamma^{(0)} &=& \exp\{-3\gamma(1-p_\gamma^{(0)})^2\}
\end{eqnarray}
The above structure is not surprising for a ferromagnetic model since 
$1-p_\gamma^{(0)}$ is nothing but
the fraction of sites which have non-vanishing field and are
therefore totally magnetized. The saddle-point equations simplify
once rewritten in terms the probability distribution $P(m)$ of the
local magnetizations $m_i=0,1$ which takes the particularly simple 
form
\begin{equation}
\label{eq:op}
P(m) = p_\gamma^{(0)} \delta_{m,0} + (1-p_\gamma^{(0)}) \delta_{m,1},
\end{equation}
where $\delta_{\cdot,\cdot}$ is the Kronecker-symbol. Thus, a fraction
$1-p_\gamma^{(0)}$ of all logical variables is {\it frozen} to +1 in all
ground states, whereas the others take both truth values with same
frequency. The self-consistent equation for $p_\gamma^{(0)}$ in
Eq.(\ref{eq:psaddle}) can be rewritten as
\begin{equation}
\label{eq:saddle}
p_\gamma^{(0)} =\sum_{c=0}^\infty
e^{-3\gamma}\frac{(3\gamma)^c}{c!}  (1-(1-p_\gamma^{(0)})^2)^c \; \;,
\end{equation}
and can be justified by simple probabilistic arguments: A variable is
frozen if and only if it is contained in at least one hyper-edge
$\{i,j,k\}\in E$ where also the two neighbors are frozen. Thus a
variable is unfrozen, $m_i=0$, if and only if every adjacent
hyper-edge contains at least one more unfrozen variable. For a spin of
connectivity $c$, this happens according to Eq.(\ref{eq:op}) with
probability $(1-(1-p_\gamma^{(0)})^2)^c$. The average over the
Poisson-distribution $e^{-3\gamma}(3\gamma)^c/c!$ of connectivities
$c$ results in the total probability for a variable to be unfrozen, so
Eq.(\ref{eq:saddle}) follows. As an additional result of replica
theory, we derive the ground-state entropy
\begin{equation}
s(\gamma) = \lim_{N\to\infty}\frac{1}{N}\ln{{\cal{N}}_{gs}} = \ln(2)
\left( p_\gamma^{(0)}(1-\ln p_\gamma^{(0)}) -\gamma[1-
(1-p_\gamma^{(0)})^3]\right) \ .
\label{eq:entropy}
\end{equation}

For small $\gamma$, Eq.(\ref{eq:saddle}) has only the trivial solution
$p_\gamma=1$ where all variables are unfrozen, i.e.\ $m_i=0$ for all
$i$.  No internal structure is found in the set of satisfying
assignments and, choosing randomly two of them, they have Hamming
distance $0.5N + O(\sqrt{N})$. To leading order in $N$, the $M$
4-clauses sets act independently, each dividing the number of
satisfying assignments by two, {\it i.e.}
${\cal{N}}_{gs}=2^{N(1-\gamma)}$. This is a clear sign that the
structure of the hyper-graph is still tree-like.

At $\gamma_d=0.818$, a new solution of Eq.(\ref{eq:saddle}) appears
discontinuously, having a fraction $(1-p_\gamma^{(0)})=0.712$ of
completely magnetized variables. This transition can be seen as a
percolation transition of fully magnetized triples of connected
variables. The entropy of this solution remains however smaller than
the entropy $1-\gamma$ of the paramagnetic solution, thus the total
solution space is still correctly described by $m_i=0$ for all
$i=1,...,N$. The appearance of the new solution signals however a
structural change in the set of solutions which breaks into an
exponential number of clusters. The cluster containing the imposed
solution $x_i= +1$ is described by the new meta-stable solution.

Another important difference to the low-$\gamma$ phase is an
exponential number of local minima of the energy function (\ref{Ham})
showing up at $\gamma_d$. These have positive energies, and the
corresponding logical assignments do not satisfy the hSAT
formula. Algorithms which decrease the energy in every time step by
local variable changes, like e.g.\ zero-temperature Glauber dynamics
or greedy algorithms, get almost surely trapped in these states and do
not find a zero-energy ground state for $\gamma>\gamma_d$. Randomized
algorithms may escape from these minima, but as found numerically,
this causes a polynomial slowing down.

By increasing $\gamma$ above $\gamma_d$, the number of ground-state
clusters decreases further. At $\gamma_c=0.918$ all but one ground-state
clusters disappear, and the non-trivial solution of (\ref{eq:saddle})
becomes the stable one. So only the cluster including the imposed
solution survives, it still contains $2^{0.082N}$ solutions, but
88.3\% of all variables are fixed to $+1$, thus forming the {\it
backbone} which appears discontinuously. As is known from
Ref.~\cite{nature}, the existence of an extensive backbone is closely
related to the exponential computational hardness of a problem.  The
remaining 11.7\% of un-frozen variables change their values from
ground-state to ground-state. They are contained in small disconnected
components or dangling ends of the hyper-graph.

The behavior of frustrated hSAT is similar, as given both by numerical
analysis and by RS or variational RSB calculations. We find that the
solution $x_i=+1$ (and its corresponding cluster) in satisfiable hSAT
are just superimposed to the solution structure of random hSAT. Thus,
the statistical properties of the solutions do not change for
$\gamma<\gamma_c$, including also the clustering of solutions above
$\gamma_d$. At $\gamma_c=0.918$ the model undergoes a SAT/UNSAT
transition, and the solution entropy jumps from 0.082 down to minus
infinity. The variational RSB calculation gives a value for the
dynamical critical connectivity $\gamma_{d,rsb}^{var}\simeq 0.83$
which is close to the exact value $0.818$. This result gives evidence
for the validity of the variational approach in the region where local
minima first appear, i.e.\ where the result does not depend strongly
on the specific functional Ansatz made for the RSB probability
distributions. For the SAT/UNSAT static transition the predictions of
the variational RSB analysis can be strongly affected by the
restriction of the functional space which does not necessarily match
the geometrical structure (clustering) of the space of
solution. However, in the case of hSAT the results are still in good
agreement, we find $\gamma_{c,rsb}^{var}\simeq 0.935$.

\section{Connection with graph theory}

In the hSAT model, we are able to extract exact results -- without the
need of RSB -- by identifying the topological structures in the
underlying hyper-graph which are responsible of the SAT/UNSAT
transition (or of frustration and glassiness). The presence (or the
absence) of such topological structures in the hyper-graph drastically
changes the statistical mechanical properties of the model.  The
different phase transitions can be viewed as different kinds of
percolation in the random graph theory language~\cite{bollobas}.

We have already seen that the formation of a locally stable
ferromagnetic state in unfrustrated hSAT at $\gamma_d=0.818$ can be
understood in term of percolation arguments. The same arguments reveal
that at $\gamma_d$ many metastables states appear in both versions of
the model, giving rise to a dynamical transition.

\begin{figure}
\centering\epsfxsize=0.6\textwidth \epsffile{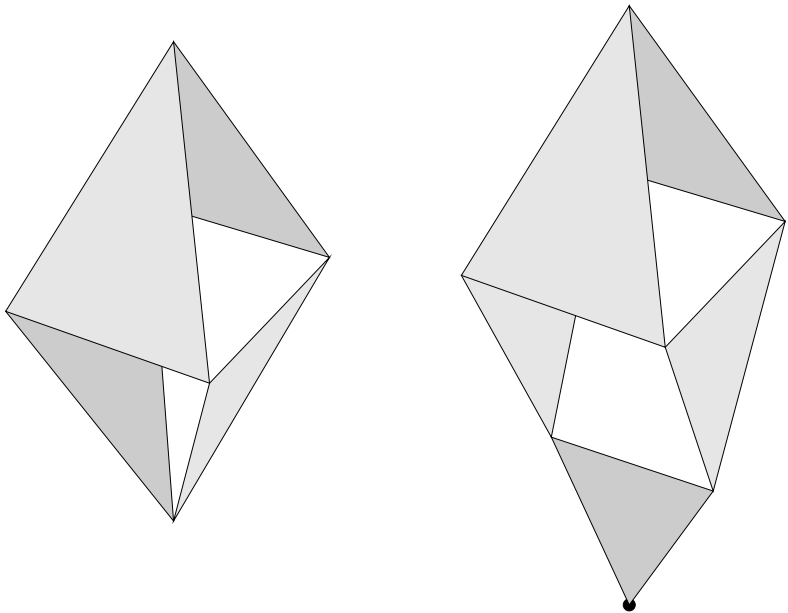}
\caption{The simplest hyper-loop (left) and the hyper-loop with one
totally constrained vertex of odd degree
(right)~\protect\cite{note5}. Triangles represent the interaction
between the three spins located at the vertices. The black dot
represents the constrained spin residing on the odd-degree vertex of
the hyper-loop.}
\label{FIG0}
\end{figure}

In order to understand what happens at the critical point $\gamma_c$
we need to introduce the notion of {\em hyper-loops}, that is the most
natural generalization of the usual {\em loop} to hyper-graphs whith
multi-vertex links.  Given a random graph ${\cal G}=(V,E)$, where $V$
is the set of vertices and $E$ is the set of (hyper-)links, a
hyper-loop can be defined as a non-zero set of (hyper-)links, $R
\subset E$, such that the degree of the subgraph ${\cal L}=(V,R)$ is
even, i.e.\ every vertex belongs to an even number of (hyper-)links
(including zero). In Fig.~\ref{FIG0} (left) we show the smallest
hyper-loop in a $K=3$ random hyper-graph. Note that in random
hyper-graph typical hyper-loops are very large and the one shown in
Fig.~\ref{FIG0} (left) is extremely rare for N large.

In a similar way we can identify those vertices which are totally
constrained. A set of (hyper-)links, $T^{(i)} \subset E$, constrains
completely the spin at site $i$ if in the sub-graph ${\cal
F}=(V,T^{(i)})$ the vertex $i$ has an odd degree and the remaining
vertices an even one.  In Fig.~\ref{FIG0} (right) we show the smallest
of such structures.

In a zero-energy configuration (SAT assignment) we have 
$S_i S_j S_k = J_{ijk}, \forall \{i,j,k\} \in E$.  Then, given any
hyper-loop $R$, we conclude
\begin{equation}
\prod_{\{i,j,k\} \in R} J_{ijk} = \prod_{\{i,j,k\} \in R} S_i S_j S_k
= 1 \quad ,
\label{hl}
\end{equation}
where the second equality comes from the fact that in the second
product every spin appears an even number of times.

In frustrated hSAT the couplings are randomly fixed to $\pm 1$ and,
consequently, the first product in Eq.(\ref{hl}) is equal to -1 with
probability $1/2$. Then we can conclude that as soon as one hyper-loop
arises in the hyper-graph half the formulas become unsatisfiable. In
general, given a hyper-graph with $N_{hl}$ hyper-loops, the fraction
of SAT formulas (with that given hyper-graph) is $2^{-N_{hl}}$.
Still one needs to average this fraction over the random hyper-graph
in order to obtain the right fraction of SAT formulas.

We have numerically found that at the critical value $\gamma_c=0.918$
the percolation of hyper-loops takes place, that is, in the large $N$
limit, the average number of hyper-loops $N_{hl}(\gamma)$ is zero for
$\gamma<\gamma_c$ and ${\cal O}(N)$ for $\gamma>\gamma_c$.  This is
the direct explanation of the SAT/UNSAT transition in terms of
hyper-graph topology.

In the unfrustrated model $J_{ijk}=1$ and Eq.(\ref{hl}) is always
satisfied. However, the mean number of hyper-loops $N_{hl}(\gamma)$ is
related to the entropy of satisfying assignments through $s(\gamma) =
(1-\gamma+N_{hl}(\gamma)/N)\ln 2$.  The derivation of this equality
straightforward if we consider the linear system modulo 2 of $M$
equations in $N$ variables, introduced at the end of
section~\ref{sec:model}.  In terms of the linear system hyper-loops
represents combination of equations giving a trivial one (e.g.\ $0=0$)
which does not fix any degree of freedom. The entropy, which is
proportional to the number of degree of freedom, is then given by
$S(\gamma) = \ln(2) (N-M-N_{hl}(\gamma))/N$.

Considering now a totally constrained spin at site $\ell$ and a SAT
assignment, we have that
\begin{equation}
\prod_{\{i,j,k\} \in T^{(\ell)}} J_{ijk} = \prod_{\{i,j,k\} \in
T^{(\ell)}} S_i S_j S_k = S_\ell \quad .
\label{hf}
\end{equation}
Then, in every SAT formula, hyper-loops with one odd-degree vertex (to
be denoted by the label $hl-1$) fix one spin variable to a complicated
function of the couplings.  We have numerically checked, that such
structures arise at $\gamma_c$, like hyper-loops, but in a
discontinuous way.

In satisfiable hSAT we have $J_{ijk}=1$, so that any
independent hyper-loop with one odd degree vertex fixes one spin to
1~\cite{note6}. Then the magnetization of the model is equal to the
mean density of such loops, $m(\gamma) = \rho_{hl-1}(\gamma) =
N_{hl-1}(\gamma)/N$. Because of the discontinuous nature of the
transition the limits $\lim_{\gamma \to \gamma_c^-}
\rho_{hl-1}(\gamma) = 0$ and $\lim_{\gamma \to \gamma_c^+}
\rho_{hl-1}(\gamma) = m_c = 0.883$ do not coincide.

In frustrated hSAT Eq.(\ref{hf}) fixes the variables belonging to
the backbone. Then one would be tempted to relate the backbone size to
the density $\rho_{hl-1}(\gamma)$ of hyper-loops with one frozen
vertex (which is true) and to estimate the backbone size at the
critical point to be 88.3\% (which is not true). Indeed at the
critical point there is a coexistence of SAT and UNSAT formulas (see
next section) and for $\gamma > \gamma_c$ all the formulas become
UNSAT in the large N limit. Then Eq.(\ref{hf}) can be applied only for
$\gamma<\gamma_c$ where the density $\rho_{hl-1}$ goes to zero when $N
\to \infty$. While the appearance of the backbone is necessarily
related to the presence of hyper-loops with frozen vertices, the
estimation of its size is non-trivial.  A very rough estimate can be
obtained assuming that at the critical point half the formulas are
SAT (according to the numerical results presented in the next section)
and that the backbone size is 0 for UNSAT formulas and 0.88 for SAT
ones. Under these very crude hypothesis the backbone size would be
0.44, which in not too far from the numerical result (see next
section).

\section{Numerical results}

We have performed extensive numerical experiments on both versions of
hSAT in order to confirm analytical predictions and to compute
quantities which are not accessible analytically.  Beside the $GF[2]$
polynomial method, we have also used two local algorithms, namely the
Davis-Putnam (DP) complete backtrack search~\cite{DP} and the
incomplete walk-SAT randomized heuristic search~\cite{wsat} , to check
the hardness of the problem for local search. The existence of at
least one solution in the satisfiable hSAT allowed us to run walk-SAT
in the whole range of $\gamma$, the halting criterion always being
finding a SAT assignment.

\begin{figure}
\centering\epsfxsize=0.8\textwidth
\epsffile{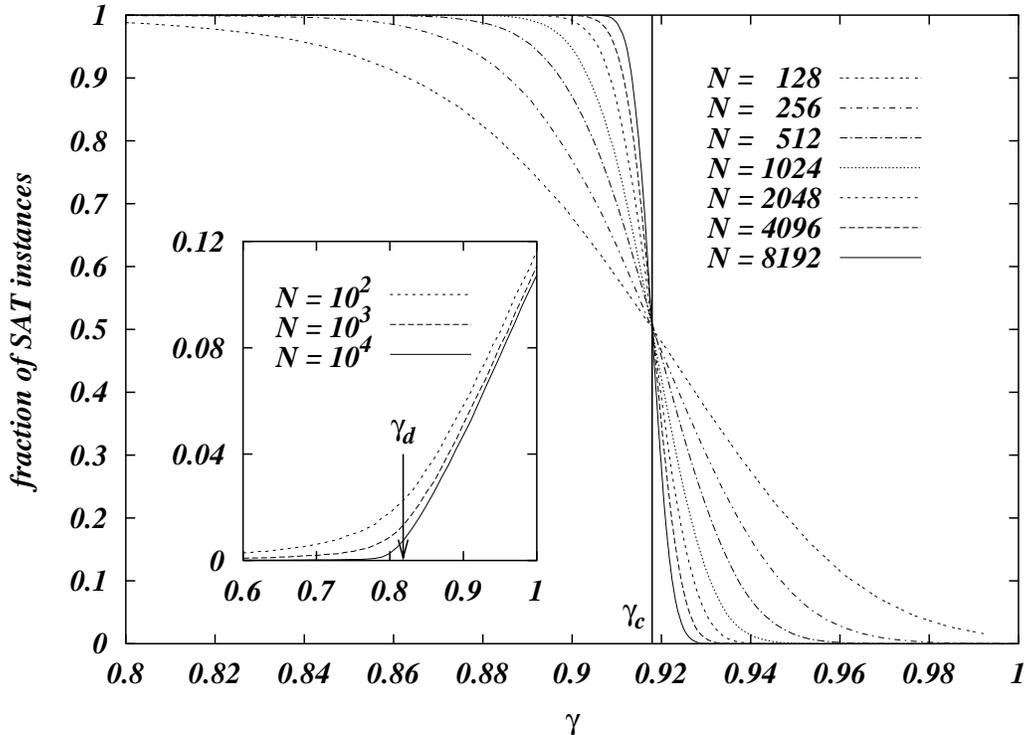}
\caption{The probability that a formula is SAT as a function of the
coupling density. Inset: The energy reached by a deterministic rule
becomes different from zero at the dynamical critical point.}
\label{FIG1}
\end{figure}

The first set of results concerns the numerical determinations of the
critical points of hSAT obtained by the polynomial method over large
samples.

For the frustrated case, the fraction of satisfiable instances drops
down to zero at $\gamma_c = 0.918$. In Fig.~\ref{FIG1} we show this
fraction as a function of $\gamma$, which has been obtained, for any
size $N$, counting the number of hyper-loops in $10^4$ different
random hyper-graphs. For any given random hyper-graph the fraction of
SAT formulas is given by $2^{-N_{hl}}$, where $N_{hl}(\gamma)$ is
the number of hyper-loops.  The same set of simulations run on the
satisfiable hSAT show that at exactly the same $\gamma_c$, the model
undergoes a discontinuous ferromagnetic transition.

At $\gamma_d = 0.818$ a dynamical transition takes place in both
versions of hSAT. There appears an exponentially large number of
positive energy local minima strongly affecting non-randomized
dynamics, which is not able to overcome energy barriers. We can easily
detect the dynamical transition by adopting the following
deterministic algorithm as a probe and by checking where it stops
converging to solutions.  The algorithm exploits the only local source
of correlations among variables that is fluctuations in
connectivity. At each step, the algorithm chooses the variable with
highest connectivity, fixes its value at random and it simplifies the
formula (``unit clause propagation''~\cite{DP}).  As can been seen in
the inset of Fig.~\ref{FIG1} the energy reached running the above
process on very large formulas ($N=10^2,10^3,10^4$) starts to deviate
from zero at a value which is highly compatible with the analytical
prediction $\gamma_d = 0.818$.  Unfortunately the mathematical
analysis of this kind of algorithm appears to be beyond our present
skills due to the correlations induced into the simplified formulas
by the particular choice of variables.  For a simple random
(connectivity independent) choice of the variable the algorithm can be
analyzed along the lines of Ref.~\cite{franco} and a convergence can
be proven up to $\gamma = 2/3$, which is also a rigorous lower bound
to the true critical density $\gamma_c$. A rigorous upper bound is
easily established by noticing that the probability for the
satisfiability of a formula at fixed $\gamma$ is bounded by the number
of satisfying assignments, averaged over all formulas of length
$\gamma N$. It follows $ \gamma_c < 2\ln 2 $ (which is the so called
{\it annealed} bound known in the statistical mechanics of disordered
media).

\begin{figure}
\centering\epsfxsize=0.8\textwidth
\epsffile{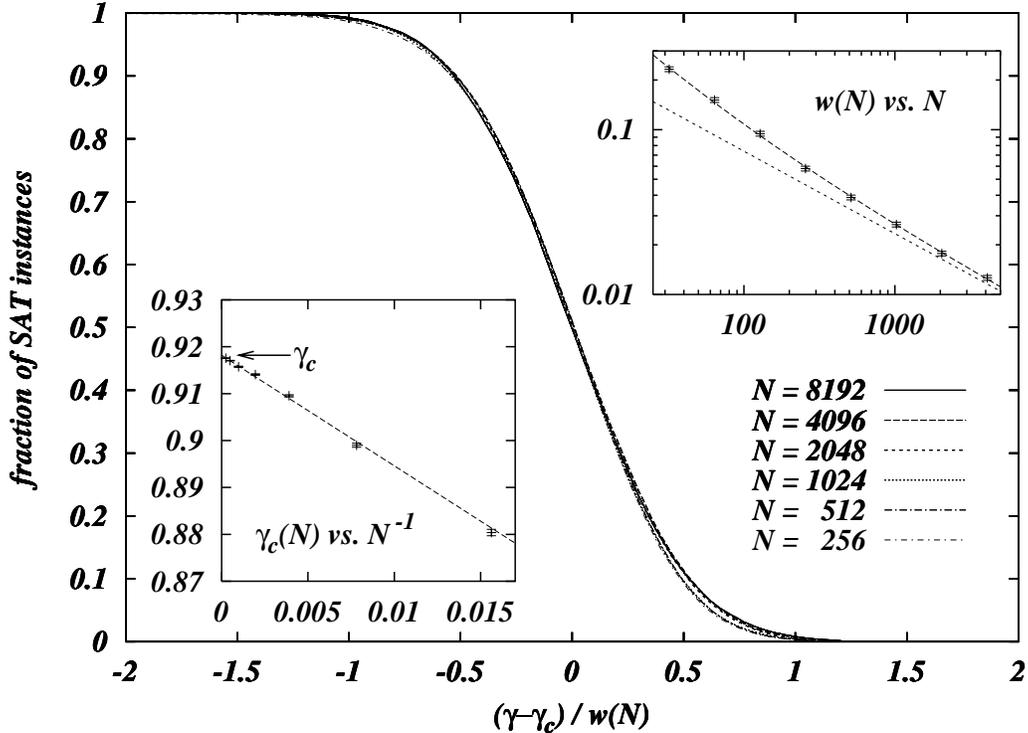}
\caption{Scaling function for the SAT probability. Lower inset: The
$\gamma$ value where the first hyper-loop arises scales as $N^{-1}$.
Upper inset: The critical width undergoes a crossover from $\nu=1$ to
$\nu=2$. The fitting curve is $3.4/N+0.74/\sqrt{N}$, while the line is
the asymptote $0.74/\sqrt{N}$.}
\label{FIG2}
\end{figure}

We have performed standard finite size scaling analysis in order to
determine the size of the critical window $w(N)$ and the $\nu$
exponent defined by $w(N) \propto N^{-1/\nu}$ for large $N$.

In a growing random hyper-graph as soon as the first hyper-loop arises
the fraction of SAT formulas drops down to 0.5. We have measured the
mean $\gamma$ value where this event takes place, $\gamma_c(N)$. Such
value scales as $\gamma_c(N) - \gamma_c \propto N^{-1}$, i.e.\ its
critical exponent is $\nu=1$ as expected for a discontinuous
transition (see lower inset in Fig.~\ref{FIG2}).

However in the model there is also another source of pure statistical
(not critical) fluctuations~\cite{wilson}. These fluctuations come
from the fact that almost every formula can be modify by the addition
(or deletion) of order $\sqrt{N}$ clauses without changing its
satisfiability.  Therefore in the large $N$ limit these purely
statistical fluctuations will dominate the critical ones, leading to
an exponent $\nu=2$ in the scaling of the SAT probability.  In the
upper inset in Fig.~\ref{FIG2} we show the width of the critical
region~\cite{note8} as a function of $N$, together with the best fit
of the kind $A x^{-B} + C x^{-1/2}$. Notably the best fitting value
for $B$ is perfectly compatible with 1, giving more evidence to the
crossover from critical fluctuations ($\nu=1$) to statistical ones
($\nu=2$).

In the main part of Fig.~\ref{FIG2} we shown the scaling function for
the SAT probability. Note that the value at criticality is equal to
0.5 up to the numerical precision.  Slight deviations from perfect
scaling appear in the $\gamma>\gamma_c$ region.  However scaling
relations hold only close to the critical point and our data perfectly
collapse in all the range where the SAT probability is between 0.2 and
0.8.

\begin{figure}
\centering\epsfxsize=0.8\textwidth
\epsffile{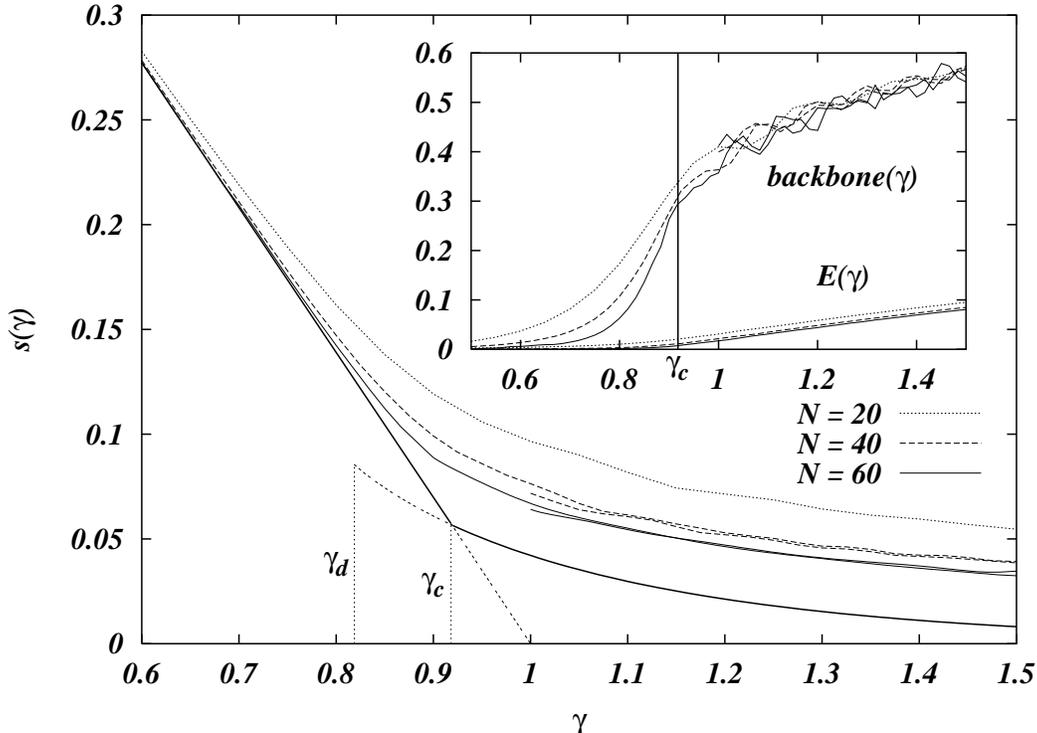}
\caption{The lowest lines are the analytical expressions for the
entropy of the unfrustrated model. The numerical estimation (not
reported) perfectly coincide.  Dashed parts correspond to metastable
states.  The rest of the data (entropy in the main body and energy and
backbone size in the inset) come from exhaustive enumeration of the
ground states in the frustrated model and of first excited states in
the unfrustrated one (only $N=40,60$) and they coincide.}
\label{FIG3}
\end{figure}

The different kind of transition taking place at $\gamma_c$ in the two
versions of hSAT is reflected in the behavior of their ground-state
entropies $s(\gamma)$ shown in Fig.~\ref{FIG3}.  For $\gamma \le
\gamma_c$ both entropies coincides and they have the analytical
expression $s(\gamma) = \ln(2) (1-\gamma)$ up to $\gamma_c$.  For
$\gamma > \gamma_c$, while the entropy of satisfiable hSAT decreases
exponentially fast with $\gamma$ (the solutions are more and more
concentrated around the superimposed one), in the frustrated version
the entropy decreases more slowly with $\gamma$, indicating that the
number of unsat assignments minimizing the energy remains
exponentially large up to $\gamma \gg \gamma_c$.

At the SAT/UNSAT transition the solution space acquires a backbone
structure, with a finite fraction of the variables that take the same
value in all the solutions.  Above the critical threshold a similar
structure characterizes the ground states. In the inset of
Fig.~\ref{FIG3} we report the results of exhaustive ground states
enumeration on small systems, giving the average size of the backbone
and the average energy per hyper-link. Increasing the system size, the
average energy converges to zero for $\gamma < \gamma_c$ and it
becomes positive at $\gamma_c$ in a continuous way.  The appearance of
the backbone becomes sharper increasing the system size and, in the
thermodynamical limit ($N \to \infty$), we expect it to be zero for
$\gamma < \gamma_c$ and finite for $\gamma \ge \gamma_c$, consistently
with a random first order phase transition predicted by the replica
theory.  As can be seen in the inset of Fig.~\ref{FIG3} the backbone
size does not depend strongly on the system size in the UNSAT phase.
As discussed in Ref.~\cite{nature} the presence of a finite backbone
is conjectured to be the source of computational hardness in finding
solutions at the SAT/UNSAT transition for both complete and randomized
local algorithms.

In the $\gamma > \gamma_c$ region the backbone size shows clear
oscillations, due to finite size effects.  At fixed energy the
backbone size is a non-decreasing function of $\gamma$, but it
typically decreases when the energy jumps to a higher value.  For
finite systems such jumps, which are of order $1/N$, are particularly
evident and induce observable fluctuations in the backbone. We expect
these fluctuation to disappear in the thermodynamic limit.

In satisfiable hSAT, once we consider only the lowest local minima
configurations just above the zero energy solutions (the so called
{\it excited states}) we find that they share completely the same
statistical properties with the ground states of the corresponding
frustrated hSAT, i.e.\ the model defined over the same random
hyper-graph with randomized couplings.  We have performed a set of DP
runs in satisfiable hSAT similar to the ones used previously, with the
additional requirement of not considering the superimposed
solution. The backbone size, the average energy and the entropy of the
excited states just above the solution are identical to those measured
on the ground states of the frustrated version (see
Fig.~\ref{FIG3}). These results, together with some preliminary
analytical findings~\cite{johannes}, show in detail why a model
without quenched frustration behave and can be modeled as having
random sign interactions, i.e.\ like a spin glass model. Such a
mapping is believed to play a particularly important role in spin
glass theory of structural glasses, in which the only source of
frustration is geometrical (i.e.\ dynamical).  Once the Boltzmann
temperature $T$ is introduced in the model, the critical points of
hSAT can be thought of as zero temperature limits of critical lines in
the $(T,\gamma)$ plane. In spite of the absence of any static
frustration and of the existence of a pure ``crystalline'' state (the
spin configurations corresponding to the satisfying assignment), the
spin system undergoes several dynamical and static transitions as the
temperature is lowered. Both the crystalline state and the first
excited states are never reached in any sub-exponential time and the
system stays for very large times in the metastable states (the same
happens in the frustrated version).

\begin{figure}
\centering\epsfxsize=0.8\textwidth
\epsffile{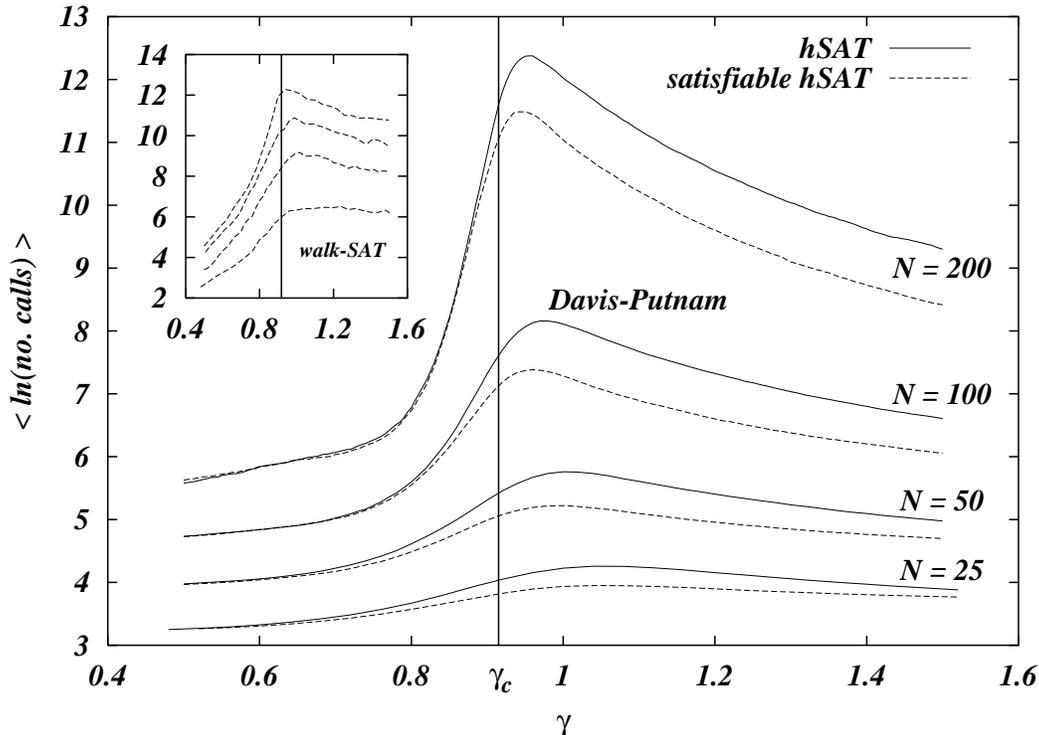}
\caption{The computational costs for finding a solution or proving
unsatisfiability with the Davis-Putnam algorithm strongly increase
approaching the critical point.  For $\gamma \ge \gamma_c$ they grow
exponentially with the problem size $N$.  Inset: The same
computational costs for the walk-SAT algorithm, which can be run for
every $\gamma$ in the satisfiable model ($N=25,50,75,100$).}
\label{FIG5}
\end{figure}

In Fig.~\ref{FIG5} we report data concerning the computational costs
for finding a solution in the satisfiable hSAT and for proving
satisfiability for the frustrated hSAT~\cite{note7}. For both
algorithms (DP and walk-SAT) and in the whole range of $\gamma$, we
have measured the logarithm of the running time averaged over
thousands of samples of different sizes.  The choice of analyzing the
averaged logarithm instead of the logarithm of the average is dictated
by the presence of fat tails in the running time distributions, even
in the $\gamma < \gamma_c$ region. The averaged logarithm provides
directly the information on the most probable cost.

The main body of the figure displays the DP computational costs for
proving satisfiability in hSAT and for finding the satisfying
assignment in the satisfiable hSAT (given the same underlying
hyper-graph structure). Both costs show a sharp easy-hard transition
at $\gamma_c$, where an enormous increase in the typical running times
take place.  For $\gamma < \gamma_c$ both costs obviously coincide and
they increase as a power law of $N$, the only effect of $\gamma_d$
being a change of the exponent from 1 to a large value which
eventually diverges at $\gamma_c$.  For $\gamma > \gamma_c$, the
computational costs remain very high, i.e.\ $<\ln[\tau(\gamma)]>
\propto \sigma(\gamma) N$, with an exponent $\sigma$ slowly decreasing
as $1/\gamma$~\cite{note10}.

In the inset of Fig.~\ref{FIG5} we show the average logarithm of the
running times needed by walk-SAT for finding a solution in the
satisfiable hSAT model.  Analogously to DP the walk-SAT costs undergo
an easy-hard transition at $\gamma_c$. Interestingly enough, above
$\gamma_c$ the computational cost for finding solutions remains quite
high and does not decrease as in DP, where the additional constraints
act as a pruning strategy in the search process.  In the hard
satisfiable region standard heuristic algorithms, like walk-SAT, get
stuck in local minima and they are not able to exploit the large
number of constraints in order to reduce the searching space.  In
particular, the large scale structure ($O(N)$) of the hyper-loops
makes them difficult to be detected in polynomial time by a local
search process which is dominated by the exponential branching process
arising at each step when the tentative choices for the variables are
made.  However, having at hand a model on which new heuristic
algorithms can be tested, such a searching optimization can be
hopefully pushed far forward.

A thorough analysis of the dependence of computational costs on 
$N$ gives the following overall picture. For $\gamma \le \gamma_d$ the
cost is a linear function of $N$. For $\gamma \in [\gamma_d,
\gamma_c)$ the typical cost increases as a power law of $N$, with an
exponent which should diverge in $\gamma_c$. For $\gamma \ge \gamma_c$
the costs are exponential in $N$.

\section{Conclusions}

In this paper we have studied a model for the generation of random
combinatorial problems, called hyper-SAT.  In the context of
theoretical computer science such a model is simply the completely
balanced version of the famous K-SAT model, while in statistical
physics it corresponds to a diluted p-spin model at zero temperature.
We have studied two version of the model, a frustrated one and an
unfrustrated one.

Increasing the density of interactions, $\gamma=M/N$, the model
undergoes two transitions. The first one is of purely dynamical nature
whereas the second one is static. Such phase transitions have a
straightforward interpretation in terms of the structure of the
underlying hyper-graphs, leading to a very simple connection between
theoretical computer science and graph theory, and statistical physics
of random systems.

The locations of phase boundaries, can be computed exactly within the
RS replica formalism, leading to $\gamma_d=0.818$ and
$\gamma_c=0.918$. We expect the replica results to be computable also
by more rigorous probabilistic methods.

Exploiting a global solution method which is polynomial in the problem
size, we have been able to study very large problems, determining with
high precision critical points and critical exponent, and a cross-over
from critical fluctuations to statistical ones has been measured.

We have found that the computational costs for finding a solution to a
typical problem or to prove that it is unsatisfiable using only local
search methods undergoes an easy-hard transitions at $\gamma_d$ and
$\gamma_c$.  The growth of the costs with the problem size $N$ is
linear up to $\gamma_d$, is polynomial in $N$ in the range $\gamma_d
\le \gamma \le \gamma_c$ and finally it becomes exponential in $N$
above $\gamma_c$. The above scenario has been checked for both
complete and incomplete local algorithms, thanks to the existence of
an halting criterion in the unfrustrated version of hyper-SAT where at
least one solution is guaranteed to exist. The use of this model as a
benchmark for heuristic algorithms may result in a good improvement of
their performances in the phase where many local minima are present.

hSAT can be viewed as an intermediate model between 2-SAT and K-SAT
($K>2$), which is exactly solvable and in which the presence of hidden
solutions can be kept under some control. Hopefully, some of the
results and methods of our analysis of hSAT can be extended to
NP-complete problems.

\section{Acknowledgements}

We thank J. Berg, S. Franz, M. Leone and D.B. Wilson for very useful
discussions. MW acknowledges the hospitality of the ICTP where the
major part of this work was done.


\begin{thebibliography}{99}
\bibitem{note2} In this context, the study of Boolean satisfiability
(K-SAT) problems has played a major role, allowing for both
theoretical and numerical analysis.  An instance of random K-SAT
consists of a set of $M=\alpha N$ random logical clauses over $N$
Boolean variables.  Each clause contains exactly $K$ variables which
are connected by logical OR operations and appear negated with
probability 1/2. The important computational question is whether there
exists an assignment to the variables that simultaneously satisfies
all clauses (``constraints'') for a given instance.  When $\alpha$
crosses a critical value $\alpha_c(K)$ and for $N\gg 1$, the
probability of finding solutions vanishes abruptly, i.e.\
$\alpha_c(K)$ identifies the so called satisfiable to unsatifiable
(SAT/UNSAT) transition.  At the same point and for $K \ge 3$, hard
computational instances cluster, leading to an exponential median
search cost for the most efficient known algorithms. The random K-SAT
problem therefore provides a good model for the study of the onset of
true intractability of ``typical'' instances of NP-complete problems.
Moreover, recents results~\cite{nature} have pointed out a clear
connection between typical-case computational complexity and the type
of threshold phenomena taking place at $\alpha_c(K)$.
\bibitem{note3} Similarly to the hSAT model, also random K-SAT can be
mapped onto a minimization problem of a spin cost function.  Keeping
the same definition of the spin variables one can construct a function
which counts the number of violated clauses.  In the $K=3$ case such a
function reads
\begin{equation}
H[\Delta ,S]= \frac{\alpha}{8} N -\sum_{i=1}^N H_i S_i+\sum_{i<j}
T_{ij} S_i S_j-\sum_{i<j<k} J_{ijk} S_i S_j S_k \; \; \;,
\label{Ham_ksat}
\end{equation}
where the couplings are given by $H_i=\frac{1}{8} \sum_{\ell=1}^M
\Delta_{\ell i}$, $T_{ij}= \frac{1}{8} \sum_{\ell=1}^M \Delta_{\ell,i}
\Delta_{\ell,j}$ and $J_{ijk}=\frac{1}{8} \sum_{\ell=1}^M
\Delta_{\ell,i} \Delta_{\ell,j} \Delta_{\ell,k}$. The random
parameters $\Delta_{\ell,i}$ take the value $1$ (resp. $-1$) if the
variable $x_i$ (resp. $\bar x_i$) appears in the $\ell$-th clause, and
$0$ otherwise.  Comparison of the above expression with Eq.(\ref{Ham})
clarifies the difference with hSAT in which only the 3-spin
interaction term appears.
\bibitem{note4} The Boolean expressions which are generated in this
way are simply those which have the highest probability given the
satisfying assignment and therefore are those with maximum entropy,
i.e.\ the easiest to satisfy.
\bibitem{note9} However the entropy is not trivially given by $\ln(2)$
times the fraction of variables which do not belong to the backbone.
\bibitem{note5} Note that in random hyper-graphs the probability that
any pair of vertices belongs to more that one hyper-link is
proportional to $1/N$ for large $N$, then we have not considered this
possibility when constructing the structures in Fig.~\ref{FIG0}.
\bibitem{note6} Taking also the gauge into account we have
$J_{ijk}=\varepsilon_i \varepsilon_j \varepsilon_k$ and the constraint
becomes $S_\ell = \varepsilon_\ell$.
\bibitem{note8} The size of the critical window has been estimated as
the inverse of the derivative at $\gamma_c$ of the curves in
Fig.~\ref{FIG1}. Other methods give compatible results.
\bibitem{note7} In any SAT phase proving satisfiability is equivalent
to finding a ground state whereas in an UNSAT phase the proof of
unsatisfiability does not provide any information on the precise
assignments of the variables in the ground states.
\end{thebibliography}
\end{document}